\documentclass[twocolumn,amsmath,amssymb,prl]{revtex4}
\usepackage{dcolumn}% Align table columns on decimal point
\usepackage{bm}% bold math
\usepackage{amsmath}
\usepackage{amssymb}
\usepackage{revsymb}
\usepackage{graphics}
\usepackage{graphicx}
\begin{document}
%\title{van der Waals forces: How do you deal with dissipative media}% Force line breaks with \\
\title{Lifshitz theory of van der Waals pressure in dissipative media}
\author{Yi Zheng}
%\email{yz2308@columbia.edu}
%\affiliation{ 
%Department of Mechanical Engineering, Columbia University\\
%New York, NY 10027
%}
\author{Arvind Narayanaswamy}
\email{arvind.narayanaswamy@columbia.edu}
\affiliation{%
Department of Mechanical Engineering, Columbia University\\
New York, NY 10027
}

\begin{abstract}
%\textbf{Needs improvement}
We derive a first--principles method of determining the van der Waals or Casimir pressure in a dissipative and dispersive planar multilayered system by calculating the Maxwell stress tensor in a fictitious layer of vacuum, that is eventually made to vanish, introduced in the structure. 
%Apart from being a general method to calculate the van der Waals energy in a dissipative medium, we present evidence for the generalization of Lifshitz theory of van der Waals forces by Dzyaloshinskii et al. without using quantum field theoretic techniques. 
This is illustrated by calculating the van der Waals pressure in a thin film with dissipative properties embedded between two semi--infinite media.
\end{abstract}
\maketitle

%Since the seminal paper by Lifshitz \cite{lifshitz56a}, van der Waals forces between macroscopic objects has been a topic of rich theoretical and experimental investigations.

%Though a variety of intermolecular interactions are important for phenomena ranging from colloidal stability to self--assembly of biomacromolecules, van der Waals interactions are common to all of them. 
 van der Waals (vdW) forces, resulting from the alteration of the quantum and thermal fluctuations of the electrodynamic field due to the presence of interfaces, play a significant role in the interactions between macroscopic objects at micrometer and nanometer length scales. Hamaker was the first to extend the concept of London--vdW forces between two atoms to forces between macroscopic spheres by pairwise summation of the interaction energy between atoms that constitute the spheres \cite{hamaker37}. 
Lifshitz, in his seminal work \cite{lifshitz56a}, outlined a method based on Rytov's theory of fluctuational electrodynamics \cite{rytov59a} for computing the vdW forces between two semi--infinite regions separated by a vacuum gap. It required the calculation of the average value of the Maxwell stress tensor in the vacuum gap.
% and the expression for vdW force involved only the electrical permittivity and magnetic permeability of the semi--infinite media. %to calculate the vdW force between the two semi--infinite objects
%While this procedure can be extended to calculating forces between objects separated by transparent media, 
The generalization of Lifshitz's method to calculating vdW forces between semi--infinite regions separated by dissipative media is not straightforward because of the difficulty in defining an electromagnetic stress tensor in dissipative media \cite{landau84a}. The goal of this paper is to obtain the general theory of vdW pressure in arbitrary planar media with dissipative and dispersive electromagnetic properties without resorting to defining the electromagnetic stress tensor or free energy in any material but vacuum.
%The goal of this paper is to use the ``simple theory'' of vdW force, as Barash refers to Lifshitz's theory \cite{barash1975a}, to obtain the general theory of vdW pressures in arbitrary planar media with dissipative and dispersive electromagnetic properties.

An approach proposed by Dzyaloshinskii, Lifshitz, and Pitaevskii (DLP from now on) \cite{dyza61a}, shrouded in the complicated language of quantum field theory, is the most frequently used generalization of Lifshitz's method  to calculate forces between objects separated by absorbing media. Even though it has been noted that an expression for Maxwell stress tensor for time--varying fields in absorbing media cannot be expressed in terms of the frequency dependent permittivity and permeability alone \cite{landau84a}, DLP method effectively reduces to using a ``Minkowski--like'' \cite{pitaevskii2006a} definition of electromagnetic stress tensor in dissipative media. Ninham et al. \cite{ninham1970a} circumvented the complications of the DLP method but, in doing so, had to postulate that the free energy of an electromagnetic mode at frequency $\omega_j$ is given by $k_BT \log\left[\sinh\left(\hbar \omega_j/2k_BT\right)\right]$, where $k_B$ is Boltzmann's constant, $2\pi\hbar$ is Planck's constant, and $T$ is the absolute temperature, even though the mode frequencies in dissipative media are, in general, complex.
%which is the expression for the free energy of an undamped harmonic oscillator at $\omega_j$. 
%The free energy for harmonic oscillators with dissipation is different. 
It has been argued by Barash and Ginzburg \cite{barash1972a,barash1975a} that ascribing to each mode a free energy of the above--mentioned form is indeed correct. The methods of DLP and Barash and Ginzburg are justified on the grounds that \emph{it is possible to ascribe thermodynamic functions to electromagnetic fields in equilibrium with matter} \cite{barash1975a,pitaevskii2006a}. 

The relative transparency of the Lifshitz method is obscured by the complexity of Dzyaloshinskii's formalism or by having to define the free energy of each mode, even though the final result is a simple generalization of the Lifshitz formula. It has been generally regarded that Lifshitz's method, in which the definition of stress tensor is above reproach, is incapable of handling dissipative media without relying on either of the two generalizations \cite{pitaevskii2006a}. 
%The agenda of this paper is to use the ``simple theory'' of vdW force, as Barash refers to Lifshitz's theory \cite{barash1975a}, to obtain the general theory of vdW pressures in arbitrary planar media with dissipative and dispersive electromagnetic properties. 
Using the fluctuation--dissipation theorem and properties of the dyadic Green's function, we express the components of the Maxwell stress tensor in vacuum in terms of components of the dyadic Green's function \cite{narayanaswamy2010a}. After a description of a general method to deal with multilayered media, we show, using examples of (1) a thin film bound by vacuum on both sides, (2) a thin film with vacuum on one side and a semi--infinite medium with arbitrary permittivity and permeability on the other, and (3) a thin film bound by semi--infinite media with arbitrary permittivity and permeability, that the expression for vdW pressure coincides with that of DLP. 

\begin{figure}
\includegraphics[width =7.0cm]{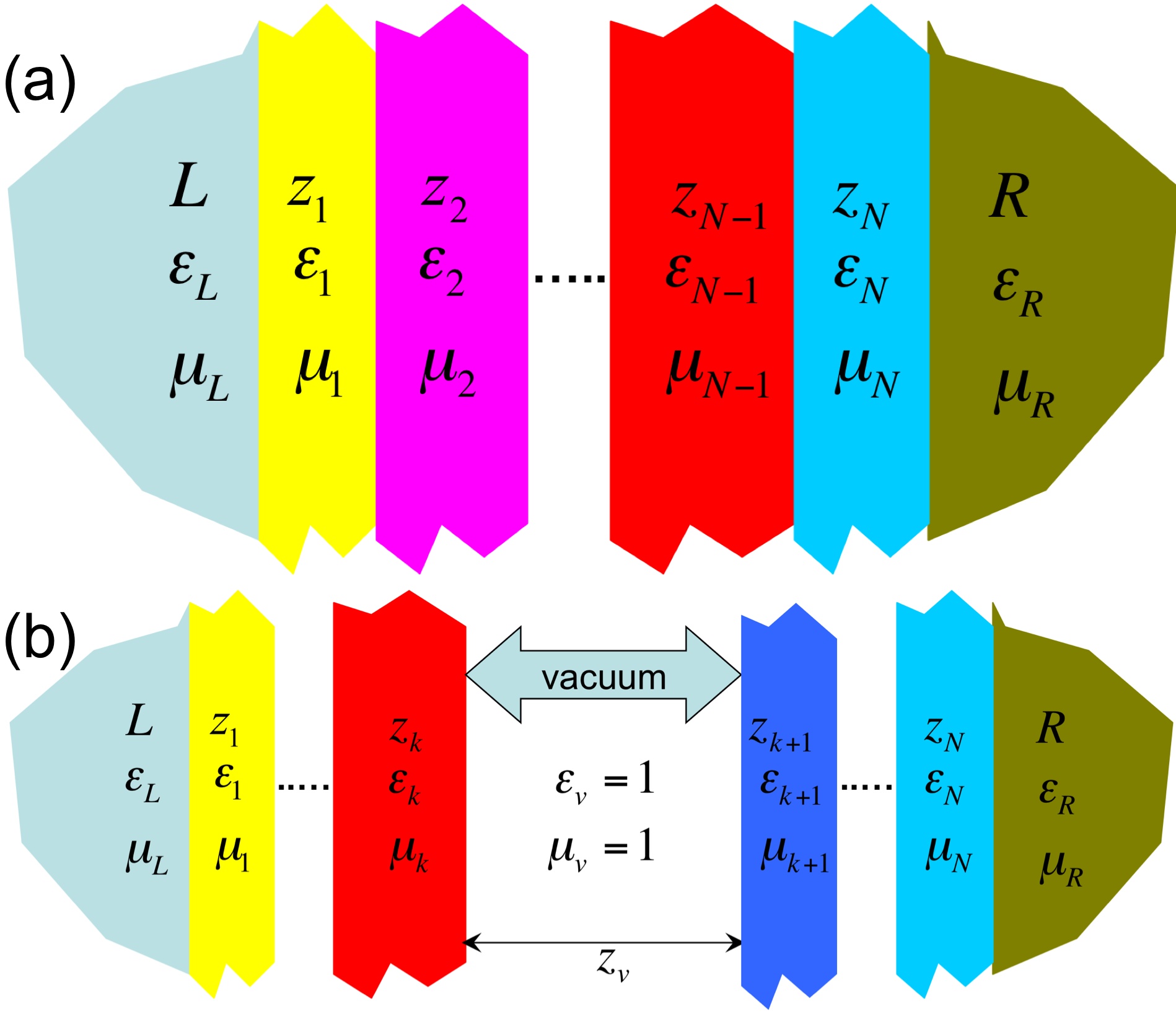}
\caption{\label{fig:vdWmethodology}(a) A multilayer system with $N$ layers between two semi--infinite regions $L$ and $R$. (b) Method of splitting $N$ layer mutilayer system into components. The $z$--axis is perpendicular to the interfaces.
%To find the vdW interaction energy of system in (a) a general multilayer system with $N$ layers sandwiched between two semi--infinite regions $L$ and $R$, we split the system into two parts as shown in (b),  consisting of $k+1$ layers bounded by $L$ to the left and vacuum to the right, and another consisting of the remaining $N-k-1$ layers with vacuum to the left and medium $R$ to the right.
}
\end{figure}

Let us analyze a general multilayer system, as shown in Fig. \ref{fig:vdWmethodology}a, and express the vdW free energy of the system in terms of combinations of vdW free energy of smaller units.
The vdW free energy per unit area of a planar configuration of $N$ layers (Fig. \ref{fig:vdWmethodology}a) sandwiched between two semi--infinite media, medium $L$ to the left and medium $R$ to the right, is represented by $U_{LR}  \left(z_1, \cdots,  z_{N}\right)$. Each layer is characterized by not only the thickness $z_k$ but also the permittivity, $\varepsilon_k$, and permeability, $\mu_k$ (both relative to that of vacuum). We use the aforementioned notation for free energy for its efficiency. If one of the semi--infinite media is vacuum, the subscript $V$ is used instead of $L$ or $R$. $U_{LR}  \left(z_1, \cdots,  z_{N}\right)$ can be written as a combination of three terms: (1) the free energy of the first $k$ layers sandwiched by semi--infinite medium $L$ to the left and vacuum to the right of the $k^{th}$ layer, $U_{LV}\left(z_1, \cdots, z_{k}\right) $, (2) the free energy of the remaining $N-k$ layers sandwiched by semi--infinite medium $R$ to the right and vacuum to the left of the $(k+1)^{th}$ layer, $U_{VR}\left(z_{k+1}, \cdots, z_{N}\right) $, and (3) the work done in bringing the two systems from infinite separation to a separation $\delta \rightarrow 0$. This statement can be written as:
\begin{equation}
\label{eqn:vdwenergygeneral}
\begin{split}
 U_{LR}  \left(z_1, \cdots,  z_{N}\right) &=  U_{LV}\left(z_1, \cdots, z_{k}\right) + \\
    U_{VR}\big(z_{k+1},&\cdots,  z_{N}\big) + \lim\limits_{\delta \rightarrow 0} \int\limits_{\infty}^{\delta} T_{zz}^{avg}\left(z_v\right) {\rm d}z_v,
  %F &\left(z_1, \cdots, z_{k}, z_v, z_{k+1},\cdots z_{N}\right) {\rm d}z_v 
\end{split}
\end{equation}
where $T_{zz}^{avg}\left(z_v\right) \equiv T_{zz}^{avg}\left(z_1,\cdots,z_k,z_v,z_{k+1},\cdots,z_N\right)$ is the vdW pressure in the vacuum region in Fig. \ref{fig:vdWmethodology}b against which work needs to be done to create the $N$ layer system from the two sub--systems. The partial derivative $\partial U_{LR}\left(z_1, \cdots, z_{N}\right)/\partial z_{r}$$=$$p^{(r)}_{LR}\left(z_1, \cdots, z_{N}\right)$ gives the vdW pressure in the $r^{th}$ layer of the $N$ layer system bounded by $L$ and $R$. For a thin film bounded by two semi--infinite regions, we drop the superscript $(r)$ and denote the pressure simply as $p_{LR}$.
By differentiating Eq. \ref{eqn:vdwenergygeneral} with respect to $z_r$, we obtain the following equation for $p^{(r)}_{LR}$:
\begin{equation}
\label{eqn:vdwpressuregeneral}
\begin{split}
 p^{(r)}_{LR}  \left(z_1, \cdots,  z_{N}\right) &=  \frac{\partial U_{LV}}{\partial z_r}\left(z_1, \cdots, z_{k}\right) + \\
    \frac{\partial U_{VR}}{\partial z_r}\big(z_{k+1},&\cdots,  z_{N}\big) +  \int\limits_{\infty}^{0} \frac{\partial T_{zz}^{avg}}{\partial z_r} \left(z_v\right) {\rm d}z_v,
  %F &\left(z_1, \cdots, z_{k}, z_v, z_{k+1},\cdots z_{N}\right) {\rm d}z_v 
\end{split}
\end{equation}
One of the first two terms on the rhs of Eq. \ref{eqn:vdwpressuregeneral} is zero depending on whether $1 \leq r \leq k$ or $k+1\leq r \leq N$.
Though $T_{zz}^{avg}\left(z_v\right)$
%$T_{zz}^{avg}\left(z_1, \cdots, z_k, z_v, z_{k+1}, \cdots,z_{N}\right) $ 
diverges as $z_v^{-3}$ for $z_v \rightarrow 0$, the quantity $\partial T_{zz}^{avg}/\partial z_{r}$ is finite as $z_v \rightarrow 0$ $\forall$ $1 \leq r \leq N$, allowing us to define the partial derivative of the last term in Eq. \ref{eqn:vdwenergygeneral} as the integral $\int\limits_{\infty}^{0} \partial T_{zz}^{avg}/\partial z_r {\rm d}z_v $ (see supplemental information for a justification).
%$T_{zz}^{avg}\left(z_1, \cdots, z_k, z_v, z_{k+1}, \cdots,z_{N}\right) $ 
$T_{zz}^{avg}\left(z_v\right) $ is obtained simply by determining the $zz$ component of the Maxwell stress tensor in the vacuum region. Using the procedure described above, we can write the vdW free energy of any $N$ layer medium in terms of $U_{VV}(z_1)$, $U_{VV}(z_2)$, $\cdots$, and $U_{VV}(z_N)$, and contributions from terms of the form $\int\limits_{\infty}^{\delta} T_{zz}^{avg}\left(z_v\right) {\rm d}z_v$, all of which involve calculation of Maxwell stress tensor in vacuum alone. %It can be seen that 
$U_{VV}(z)$ is nothing but the vdW free energy to create a thin film of thickness $z$ in free space.
%This method is not unlike that of determining the electrostatic energy of an assembly of $N$ charges 

We rely on Rytov's theory of fluctuational electrodynamics to determine the value of $T_{zz}^{avg}$.
 The cross--spectral correlations of the electric field components can be written as $\langle$$E_{p}(\bm{r},\omega)$$E^*_{q}(\bm{r},\omega)$$\rangle$$=$$(2\omega \mu_o \Theta/\pi)$$\text{Im}$$\bm{G}^{e}_{pq}(\bm{r},\bm{r}) $  ($p$,$q$=$x$,$y$,$z$),  where $\Theta$$=$$(\hbar \omega/2)$$\coth(\hbar \omega/2k_B T)$, and $\bm{G}^e_{pq}$ is the $pq$ component of the electric dyadic Green's function \cite{joulain03a,narayanaswamy2010a}. 
 The spectral correlation is defined such that $\langle$$E_{p}(\bm{r},t)$$E_{q}$$(\bm{r}$,$t$$)$$\rangle$$=$$\int\limits_{0}^{\infty}$$d\omega$ $\langle$$E_{p}(\bm{r},\omega)$$E^*_{q}(\bm{r},\omega)$$\rangle$. Similarly, the cross--spectral correlations of the magnetic field components are given by $\langle$$H_{p}$$(\bm{r}$,$\omega)$$H^*_{q}$$(\bm{r}$,$\omega)$$\rangle$$= $$(2\omega \epsilon_o \Theta/\pi)$$\text{Im}$$\bm{G}^{m}_{pq}$$(\bm{r}$,$\bm{r})$. The dyadic Green's functions $\bm{G}^e$ and $\bm{G}^m$ are electromagnetic duals of each other and are solutions of $\nabla$$\times$$\nabla$$\times$$\bm{G}(\bm{r},\bm{r'})$$-$$k^2$$\bm{G}(\bm{r},\bm{r'})$$= $$\bm{I}\delta$$\left(\bm{r}-\bm{r'}\right)$, where $\bm{I}$ is the identity dyad. $\bm{G}^e$ and $\bm{G}^m$ are obtained by enforcing the continuity of: (1) $\mu(\bm{r})$$(\bm{\hat{n}}$$\times$$\bm{G}^e\left(\bm{r},\bm{r'}\right))$, (2) $\bm{\hat{n}}$$\times$$\nabla$$\times$$\bm{G}^e\left(\bm{r},\bm{r'}\right)$, (3) $\varepsilon(\bm{r})$$(\bm{\hat{n}}$$\times \bm{G}^m\left(\bm{r},\bm{r'}\right))$, and (4) $\bm{\hat{n}}$$\times$$\nabla$$\times$$\bm{G}^m\left(\bm{r},\bm{r'}\right)$ on either side of an interface defined by the unit normal vector $\bm{\hat{n}}$ at the point $\bm{r}$. 
%\begin{equation}
%\label{eqn:electricfieldcrosscorrelation}
%\frac{\varepsilon_o}{2} \langle E_{p}(\bm{r},\omega)E^*_{q}(\bm{r},\omega) \rangle = \frac{\omega \theta(\omega,T)}{\pi c^2} \Im \left(\bm{G}^{sc}_{epq}(\bm{r},\bm{r}) \right)
%\end{equation}
%\begin{equation}
%\label{eqn:magneticfieldcrosscorrelation}
%\frac{\mu_o}{2}\langle H_{p}(\bm{r},\omega)H^*_{q}(\bm{r},\omega) \rangle = \frac{\omega \theta(\omega,T)}{\pi c^2} \Im \left(\bm{G}^{sc}_{mpq}(\bm{r},\bm{r}) \right)
%\end{equation}

\begin{figure}
\includegraphics[width=8cm]{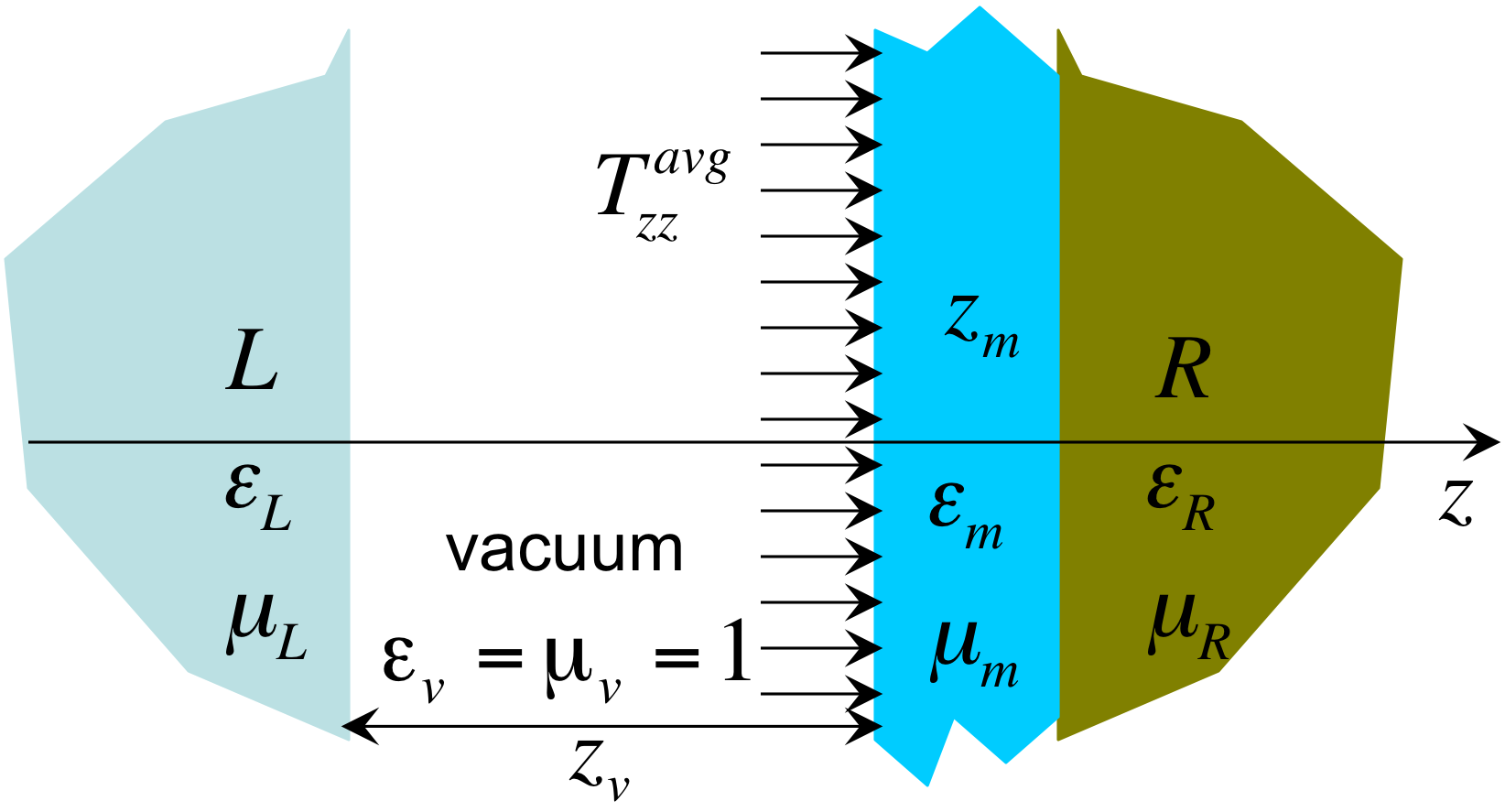}
\caption{\label{fig:4layermedium}A four layer system. 
%The stress tensor $T_{zz}^{avg}$ is a function of $z_m$ and $z_v$ and can be written in terms of the coefficients of forward and backward travelling plane waves that comprise the dyadic Green's function.
}
\end{figure}

The $zz$ component of the Maxwell stress tensor in vacuum can be expressed in terms of $\bm{G}^e$ and $\bm{G}^m$ as ${T}_{zz}(\bm{r},\omega)$$=$$(2\omega\Theta/\pi c^2)$$\text{Im}$$G\left(\bm{r},\omega\right)$ \cite{narayanaswamy2010a,joulain03a}, where $G\left(\bm{r},\omega\right)$$=$$G^{e}_{zz}(\bm{r,r})$$-\frac{1}{2}\text{Tr}\bm{G}^{e}(\bm{r,r})$$ + G^{m}_{zz}(\bm{r,r})$$-\frac{1}{2}\text{Tr}\bm{G}^{m}(\bm{r,r}) $. The average value of the $zz$ component of the Maxwell stress tensor, $T_{zz}^{avg}$, at any instant of time at position $\bm{r}$ in a vacuum layer is given by:
\begin{equation}
T_{zz}^{avg} =  \int\limits_0^{\infty} \frac{\hbar \omega^2}{\pi c^2} \coth\left(\frac{\hbar \omega}{2k_B T}\right) \text{Im}G\left(\bm{r},\omega\right) {\rm d}\omega 
\end{equation}
$\bm{G}^{e}$ and $\bm{G}^{m}$ are analytic in the upper half plane (UHP) by virtue of being response functions. Since $G(\bm{r},\omega)$ is a linear combination of different components of $\bm{G}^{e}$ and $\bm{G}^{m}$, it is also analytic in the UHP. We can therefore use Lifshitz's technique to replace the integral over $\omega$ along the real positive frequency axis  by a summation over Matsubara frequencies on the imaginary frequency axis in the UHP as:
\begin{equation}
\label{eqn:avgstresstensorsum}
\begin{split}
T_{zz}^{avg} &= 
%\frac{\hbar}{2\pi c^2} \int\limits_0^{\infty} \omega^2 \coth(\frac{ \hbar \omega}{2k_BT}) \text{Im} G\left(\bm{r},\omega\right)  {\rm d}\omega \\
%&=
-\frac{2k_BT}{ c^2}\sum _{n=0}^{\infty}{'}\xi_n^2G\left(\bm{r},i\xi_n\right)=k_BT\sum_{n=0}^{\infty}{'}K_n
\end{split}
\end{equation}
where, $\xi_n$$=$$2\pi n k_B T/\hbar$, $K_n$$=$$-2\xi_n^2$$G(\bm {r},i\xi_n)/c^2$, $K_0$$=-$$\lim\limits_{\xi \rightarrow 0}$$2\xi^2$$G(\bm {r},i\xi)/c^2$, and $n $$=$$0,1,2,$$\cdots$. The prime ($'$) next to $\sum$ indicates that the $n=0$ term is given weight 0.5. $G\left(\bm{r},i\xi_n\right)$ can be written in terms of the reflection coefficients of plane waves that comprise $\bm{G}^e$ and $\bm{G}^m$ \cite{parsegian2005a}. We now apply this method to calculating the vdW pressure in a thin film (indicated by $m$)  bounded by two semi--infinite objects, $L$ and $R$. To do so, we introduce a vacuum layer, shown in Fig. \ref{fig:4layermedium}, in which the Maxwell stress tensor will be determined. %= (i\xi_n/c)^2(\beta\pi k_v^2)^{-1}\int\limits_0^{\infty} \sum_{p=TE,TM} A^{(p)}_v k_{zv}k_{\rho} {\rm d}k_{\rho} $.

We start with the assertion that the vdW pressure in any infinite or semi--infinite planar medium is zero.  We will show using the following three examples that the proposed method is in agreement with the predictions of DLP theory for the case of a thin film between two semi--infinite objects. $K_n$, from which $T_{zz}^{avg}$ can be calculated using Eq. \ref {eqn:avgstresstensorsum}, for the configuration shown in Fig. \ref{fig:4layermedium} is given by:
%Since determining the vdW pressure in layer $m$ involves calculation of $T_{zz}^{avg}$ in the vacuum layer in Fig. \ref{fig:4layermedium}, we give an explicit expression for $K_n$ from which $T_{zz}^{avg}$ can be calculated using Eq. \ref {eqn:avgstresstensorsum}
\begin{equation}
K_n=\frac{1}{\pi}\int\limits_0^{\infty}   \sum_{p=e,h} \frac{R_{vL}^{(p)}\widetilde{R}_{vR}^{(p)}e^{-2k_{zv}z_v}}{1-R_{vL}^{(p)}\widetilde{R}_{vR}^{(p)}e^{-2k_{zv}z_v}}   k_{zv} k_\rho {\rm d}k_\rho
\end{equation}
where $p=e,h$ refer to the transverse electric and transverse magnetic polarizations respectively, and 
\begin{subequations}
\begin{eqnarray}
\label{eqn:rtildeteexpr} \widetilde{R}_{vR}^{(p)} & = & \frac{R_{vm}^{(p)}+R_{mR}^{(p)}e^{-2k_{zm}z_m}}{1+R_{vm}^{(p)}R_{mR}^{(p)}e^{-2k_{zm}z_m}}, \\
\label{eqn:rteexpr}  R_{vL}^{(e)} & = & \frac{k_{zv}\mu_L-k_{zL}\mu_v}{k_{zv}\mu_L+k_{zL}\mu_v}, R_{vL}^{(h)}  =  \frac{k_{zv}\varepsilon_L-k_{zL}\varepsilon_v}{k_{zv}\varepsilon_L+k_{zL}\varepsilon_v},\\
\label{eqn:kzvkzm} k_{zv} &=& \sqrt{k_{\rho}^2+\varepsilon_v \xi_n^2/c^2}, k_{zm}=\sqrt{k_{\rho}^2+\varepsilon_m \xi_n^2/c^2} ,
%\label{eqn:kzm} k_{zm}&=&\sqrt{k_{\rho}^2+\frac{\xi_n^2}{c^2}\varepsilon_m} \\
\end{eqnarray}
\end{subequations}
and similarly for reflection coefficients at other interfaces and wavevectors in other layers. All permittivities and permeabilities are evaluated at $i\xi_n$, $n=0,1,2,$$\cdots$. For reflection coefficients, the subscript $v$ will be used to denote an interface with vacuum. Since it is $\frac{\partial}{\partial z_m} \int_{\infty}^0 T_{zz}^{avg}(z_m,z_v){\rm d}z_v$ that will eventually be used in calculating vdW pressure, we give below the expression for $\int_{\infty}^0 \frac{\partial K_n}{\partial z_m}  {\rm d}z_v$

\begin{widetext}
\begin{equation}
\label{eqn:vdwpressureeqn}
\begin{split}
\int_{\infty}^0 \frac{\partial K_n}{\partial z_m}  {\rm d} z_v
%&=\frac{\partial}{\partial z_m} \lim\limits_{\delta \rightarrow 0} \int\limits_{\infty}^{\delta}F(\xi_n,z_v) {\rm d}z_v=\frac{\partial}{\partial z_m} \lim\limits_{\delta \rightarrow 0} \left( \frac{1}{2\beta\pi}\int\limits_0^{\infty}\sum_{p=TE,TM}\ln(1-R_{vL}^{(p)}\widetilde{R}_{vm}^{(p)}e^{-2k_\rho\delta}) k_\rho {\rm d}k_\rho \right)\\
%&=\frac{1}{\beta\pi}\int\limits_0^{\infty} \sum_{p=TE,TM}\frac{R_{vL}^{(p)}R_{mR}^{(p)}(1-R_{vm}^{(p)2})e^{-2k_{zm}z_m}}{(1-R_{vL}^{(p)}\widetilde{R}_{vm}^{(p)})(1+R_{vm}^{(p)}R_{mR}^{(p)}e^{-2k_{zm}z_m})^2}k_{zm}k_\rho {\rm d}k_\rho\\
%& 
=\frac{1}{\pi}\int\limits_0^{\infty}\sum_{p=e,h}\frac{R_{vL}^{(p)}R_{mR}^{(p)}(1-R_{vm}^{(p)2})e^{-2k_{zm}z_m}}{[(1-R_{vL}^{(p)}R_{vm}^{(p)})-(R_{vL}^{(p)}R_{mR}^{(p)}+R_{mv}^{(p)}R_{mR}^{(p)})e^{-2k_{zm}z_m}](1+R_{vm}^{(p)}R_{mR}^{(p)}e^{-2k_{zm}z_m})}k_{zm}k_\rho {\rm d}k_\rho
\end{split}
\end{equation}
\end{widetext}

%\textit{Example 1: Vacuum--Thin Film--Vacuum} Though we are interested in finding the Hamaker function for a thin film of water surrounded by vacuum, we will apply the method described above using $L \rightarrow$ water, $R \rightarrow$ vacuum, $N=0$, and only layer ($N+1$) being water. To determine the Hamaker function of a water film of thickness $z$, we find the vdW energy required to create the thin film of water at $x=\infty$ and bring it to a distance $x=\delta$ from the surface of medium $L$. 

\textit{Example 1: Vacuum--Thin Film--Vacuum} -- To find the vdW pressure in a thin film of material of thickness $z_m$, we consider a four layer configuration, as shown in Fig. \ref{fig:4layermedium}, with $L$ being replaced with material $m$, and $R$ being vacuum. If the vdW energy for creating a film of thickness $z_m$ is $U_{VV}(z_m)$, the following equation can be written for conservation of energy for moving the thin film from $z_v=\infty$ to $z_v=\delta\rightarrow0$:
\begin{equation}
\label{conservationofenergy}
U_{VV}(z_m)+\lim\limits_{\delta\rightarrow0}\int\limits_{\infty}^{\delta}T^{avg}_{zz}{\rm d}z_v=U_o,
\end{equation}
where, $U_o$ is an arbitrary constant that is the energy per unit area of a semi--infinite medium $M$ adjacent to a semi--infinite region of vacuum. %We can show that $U_o(\delta) \sim \delta^{-2} $ as $\delta/z_m \rightarrow0$ (this can be seen from the behavior of the vdW energy of a thin--film separating two semi--infinite regions in the non--retarded limit) and is independent of $z_m$. 
Differentiation Eq. \ref{conservationofenergy} with respect to $z_m$ gives the following equation for vdW pressure:
\begin{equation}
\label{eqn:vmvpressure}
p_{VV}(z_m)+\int\limits_{\infty}^{0}\frac{\partial T_{zz}^{avg}}{\partial z_m}{\rm d}z_v=0
\end{equation}
Using Eq. \ref{eqn:avgstresstensorsum}, Eq. \ref{eqn:vdwpressureeqn}, and Eq. \ref{eqn:vmvpressure}, the vdW pressure in a thin film of medium $m$ bounded by vacuum is given by (see supplemental information for further details):
\begin{equation}
\begin{split}
\label{eqn:pressureVV} p_{VV}(z_m)
%&=-\lim\limits_{\delta\rightarrow0}\frac{\partial}{\partial z_m}\int\limits_{\infty}^{\delta}T_{zz}(z_m,z_v){\rm d}z_v\\
=\frac{k_BT}{\pi}\sum _{n=0}^{\infty}{'} \int\limits_0^{\infty}\sum_{p=e,h} & \frac{R_{mv}^{(p)2}e^{-2k_{zm}z_m}}{1-R_{mv}^{(p)2}e^{-2k_{zm}z_m}} \\
& \times k_{zm}k_\rho {\rm d}k_\rho
\end{split}
\end{equation}
where each integral is evaluated at the Matsubara frequency $\xi_n = 2\pi n k_B T/\hbar$.
%\begin{figure}
%\includegraphics[width=8cm]{VacuumWaterVacuum6curves1029.png}
%\includegraphics[width=8cm]{PSWaterVacuum6curves1029.png}
%\includegraphics[width=8cm]{MWaterMGingell8curves1028.png}
%\caption{Hamaker functions,$A=\frac{F_{vdW}}{6\pi z_m^3}$, as a function of thickness of water $z_m$ for Vacuum-Water-Vacuum (Example 1), Polystryrene-Water-Vacuum (Example 2) and Plastic-Water-Plastic (Example 3). The curves generated by the proposed free-energy method are matching x markers generated by Lifshitz theory of vdW by Dyzaloshinskii et al.
%}
%\end{figure}

\textit{Example 2: Material--Thin Film--Vacuum} -- To find $p_{LV}(z_m)$, we consider a four layer configuration, as shown in Fig. \ref{fig:4layermedium}, with $R$ being vacuum. Equation \ref{eqn:vdwenergygeneral} can be modified for the four layer system to give the following equation for $U_{LV}(z_m)$:
\begin{equation}
\begin{split}
\label{eqn:ex2aenergy} U_{LV}(z_m) &=U_{VV}(z_m)+\lim\limits_{\delta \rightarrow 0} \int\limits_{\infty}^{\delta} T_{zz}^{avg}{\rm d}z_v,
\end{split}
\end{equation}
%where $U_1(\delta)$, the vdW free energy of the four layer system $L-v-m-V$ as $\delta/z_m \rightarrow 0$, is independent of $z_m$. 
Differentiating Eq. \ref{eqn:ex2aenergy} with respect to $z_m$, we obtain the following equation for $p_{LV}(z_m)$ in terms of $p_{VV}(z_m)$, which has been calculated earlier, and $T_{zz}^{avg}$:
\begin{equation}
\begin{split}
p_{LV}(z_m)&=p_{VV}(z_m)+\int\limits_{\infty}^{0} \frac{\partial T_{zz}^{avg}}{\partial z_m} {\rm d}z_v
\end{split}
\end{equation}
Using the expressions for $p_{VV}$ (Eq. \ref{eqn:pressureVV}) and 
%$\int\limits_{\infty}^{0} \frac{\partial T_{zz}^{avg}}{\partial z_m} {\rm d}z_v$ 
Eq. \ref{eqn:vdwpressureeqn}, we obtain the following equation for $p_{LV}(z_m)$ (see supplemental information for further details):
\begin{equation}
\label{eqn:pressureLV}
\begin{split}
 p_{LV}(z_m)&=
 %p_{VV}(z_m)+\int\limits_{\infty}^{0} \frac{\partial T_{zz}^{avg}}{\partial z_m} {\rm d}z_v = 
 \frac{k_B T}{\pi} \times\\
\sum _{n=0}^{\infty}{'}& \int\limits_0^{\infty}\sum_{p=e,h} \frac{R_{mL}^{(p)}R_{mv}^{(p)}e^{-2k_{zm}z_m}}{1-R_{mL}^{(p)}R_{mv}^{(p)}e^{-2k_{zm}z_m}} k_{zm}k_\rho {\rm d}k_\rho
\end{split}
\end{equation}
We can obtain the vdW pressure $p_{VR}(z_m)$ by replacing $L$ with $R$ in Eq. \ref{eqn:pressureLV}.
%\begin{equation}
%\label{ex2b} p_{vwR}=\frac{1}{\beta\pi}\int\limits_0^{\infty}\sum_{p=TE,TM} \frac{R_{mv}^{(p)}R_{mR}^{(p)}e^{-2k_{zw}z_m}}{1-R_{mv}^{(p)}R_{mR}^{(p)}e^{-2k_{zw}z_m}} k_{zw}k_\rho {\rm d}k_\rho
%\end{equation}

\textit{Example 3: Material--Thin Film--Material} -- The vdW free energy of the system $L-m-R$ is obtained by adding to the free energy $U_{VR}(z_m)$ the work done in moving this system from infinite separation to the surface of a semi--infinite region of material $L$. Written as an equation, we get:
%of a thin film of material $m$ bounder by vacuum to the left and material $R$ to the right 
\begin{equation}
\label{eqn:energyLR}
\begin{split}
U_{LR}(z_m) = U_{VR}(z_m) + \lim\limits_{\delta \rightarrow 0} \int\limits_{\infty}^{\delta} T_{zz}^{avg}{\rm d}z_v,
\end{split}
\end{equation}
and $p_{LR}(z_m)$ is given by (see supplemental information for further details):
\begin{equation}
\begin{split}
\label{eqn:pressureLR} p_{LR}(z_m)&=p_{VR}(z_m)+\int\limits_{\infty}^{0}\frac{\partial T^{avg}_{zz}}{\partial z_m}{\rm d}z_v=\frac{k_B T}{  \pi}\times\\
\sum _{n=0}^{\infty}{'}&\int\limits_0^{\infty}\sum_{p=e,h} \frac{R_{mL}^{(p)}R_{mR}^{(p)}e^{-2k_{zm}z_m}}{1-R_{mL}^{(p)}R_{mR}^{(p)}e^{-2k_{zm}z_m}} k_{zm}k_\rho {\rm d}k_\rho
\end{split}
\end{equation}
It can be seen that Eq. \ref{eqn:pressureLR} for $p_{LR}(z_m)$ is a generalization of Eq. \ref{eqn:pressureVV} and Eq. \ref{eqn:pressureLV}. Further simplification of Eq. \ref{eqn:pressureLR}, as shown in the supplemental information, results in the following expression for $p_{LR}$:
\begin{equation}
\label{eqn:DLP}
\begin{split}
 p_{LR}(z_m) =  & \frac{k_BT}{\pi c^3} \sum _{n=0}^{\infty}{'}  \varepsilon_m^{3/2} \xi_n^3\int\limits_{1}^{\infty} {\rm d}q q^2 \times \\
\sum_{p=e,h}&(R_{mL}^{(p)-1}R_{mR}^{(p)-1}e^{2q\xi_n\sqrt{\varepsilon_m}z_m/c}-1)^{-1},
\end{split}
\end{equation}
where $q = k_{zm}/(\xi_n\sqrt{\varepsilon_m(i\xi_n)}/c)$. 
%where the $n=0$ contribution is obtained by a limiting procedure described earlier. Equation \ref{eqn:DLP} 
Equation \ref{eqn:DLP} agrees with the expression for vdW pressure in a thin film according to DLP \cite{dyza61a, gingell1973a}. We stress that the method outlined here for calculating vdW pressure is valid irrespective of computation of the electromagnetic stress tensor by a summation along the imaginary frequency axis or along the real frequency axis. The extension to a multilayered medium is simply an exercise in determining the appropriate reflection and transmission coefficients \cite{parsegian2005a,chew95a}.

%van derWaals forces play an integral role in a variety of biological phenomena, colloidal stability, adhesion. In many of these cases, it becomes necessary to calculate vdW forces in aqueous media. 
%\textbf{Conclusion needs improvement} 
We have provided here a transparent formalism for calculating vdW or Casimir pressure in dissipative and dispersive media that are constituents of planar multilayer structures without having to define or calculate the stress tensor in such layers. 
%We have shown that Lifshitz theory of vdW forces is general enough to encompass dissipative and dispersive materials.
We provide evidence backing the generalization of Lifshitz theory of vdW forces without relying on quantum field theoretic techniques employed by Dzyaloshinskii, Lifshitz, and Piatevskii. These results offer further proof of the validity of the Minkowski--like stress tensor for calculating vdW forces, at least in planar multilayered media. This formalism can be generalized to obtain the vdW free energy and pressure of systems involving finite sized objects.

The authors would like to acknowledge correspondence and discussions with Prof. Adrian Parsegian, Prof. Rudi Podgornik, and Prof. L. Pitaevskii. It was brought to our attention that Pitaevskii and Lifshitz had started to solve the problem of vdW pressure in dissipative media by the method outlined in this paper but put an end to it on the suggestion of Landau\cite{pitaevskiiCasimirPhysics}!

%\bibliography{../arvindmac}

\begin{thebibliography}{15}
\expandafter\ifx\csname natexlab\endcsname\relax\def\natexlab#1{#1}\fi
\expandafter\ifx\csname bibnamefont\endcsname\relax
  \def\bibnamefont#1{#1}\fi
\expandafter\ifx\csname bibfnamefont\endcsname\relax
  \def\bibfnamefont#1{#1}\fi
\expandafter\ifx\csname citenamefont\endcsname\relax
  \def\citenamefont#1{#1}\fi
\expandafter\ifx\csname url\endcsname\relax
  \def\url#1{\texttt{#1}}\fi
\expandafter\ifx\csname urlprefix\endcsname\relax\def\urlprefix{URL }\fi
\providecommand{\bibinfo}[2]{#2}
\providecommand{\eprint}[2][]{\url{#2}}

\bibitem[{\citenamefont{Hamaker}(1937)}]{hamaker37}
\bibinfo{author}{\bibfnamefont{H.~C.} \bibnamefont{Hamaker}},
  \bibinfo{journal}{Physica} \textbf{\bibinfo{volume}{4}},
  \bibinfo{pages}{1058} (\bibinfo{year}{1937}).

\bibitem[{\citenamefont{Lifshitz}(1956)}]{lifshitz56a}
\bibinfo{author}{\bibfnamefont{E.~M.} \bibnamefont{Lifshitz}},
  \bibinfo{journal}{Sov. Phys. JETP} \textbf{\bibinfo{volume}{2}},
  \bibinfo{pages}{73} (\bibinfo{year}{1956}).

\bibitem[{\citenamefont{Rytov}(1959)}]{rytov59a}
\bibinfo{author}{\bibfnamefont{S.~M.} \bibnamefont{Rytov}},
  \emph{\bibinfo{title}{Theory of Electric Fluctuations and Thermal Radiation}}
  (\bibinfo{publisher}{Air Force Cambridge Research Center},
  \bibinfo{address}{Bedford, MA}, \bibinfo{year}{1959}).

\bibitem[{\citenamefont{Landau et~al.}(1984)\citenamefont{Landau, Lifshitz, and
  Pitaevskii}}]{landau84a}
\bibinfo{author}{\bibfnamefont{L.~D.} \bibnamefont{Landau}},
  \bibinfo{author}{\bibfnamefont{E.~M.} \bibnamefont{Lifshitz}},
  \bibnamefont{and} \bibinfo{author}{\bibfnamefont{L.~P.}
  \bibnamefont{Pitaevskii}}, \emph{\bibinfo{title}{Electrodynamics of
  Continuous Media}} (\bibinfo{publisher}{Pergamon Press},
  \bibinfo{year}{1984}).

\bibitem[{\citenamefont{Dzyaloshinskii
  et~al.}(1961)\citenamefont{Dzyaloshinskii, Lifshitz, and
  Pitaevskii}}]{dyza61a}
\bibinfo{author}{\bibfnamefont{I.~E.} \bibnamefont{Dzyaloshinskii}},
  \bibinfo{author}{\bibfnamefont{E.~M.} \bibnamefont{Lifshitz}},
  \bibnamefont{and} \bibinfo{author}{\bibfnamefont{L.~P.}
  \bibnamefont{Pitaevskii}}, \bibinfo{journal}{Adv. Phys.}
  \textbf{\bibinfo{volume}{10}}, \bibinfo{pages}{165} (\bibinfo{year}{1961}).

\bibitem[{\citenamefont{Pitaevskii}(2006)}]{pitaevskii2006a}
\bibinfo{author}{\bibfnamefont{L.}~\bibnamefont{Pitaevskii}},
  \bibinfo{journal}{Phys. Rev. A} \textbf{\bibinfo{volume}{73}},
  \bibinfo{pages}{47801} (\bibinfo{year}{2006}).

\bibitem[{\citenamefont{Ninham et~al.}(1970)\citenamefont{Ninham, Parsegian,
  and Weiss}}]{ninham1970a}
\bibinfo{author}{\bibfnamefont{B.~W.} \bibnamefont{Ninham}},
  \bibinfo{author}{\bibfnamefont{V.~A.} \bibnamefont{Parsegian}},
  \bibnamefont{and} \bibinfo{author}{\bibfnamefont{G.~H.} \bibnamefont{Weiss}},
  \bibinfo{journal}{J. Stat. Phys.} \textbf{\bibinfo{volume}{2}},
  \bibinfo{pages}{323} (\bibinfo{year}{1970}).

\bibitem[{\citenamefont{Barash and Ginzburg}(1972)}]{barash1972a}
\bibinfo{author}{\bibfnamefont{Y.~S.} \bibnamefont{Barash}} \bibnamefont{and}
  \bibinfo{author}{\bibfnamefont{V.~L.} \bibnamefont{Ginzburg}},
  \bibinfo{journal}{Sov. Phys. JETP Lett.} \textbf{\bibinfo{volume}{15}},
  \bibinfo{pages}{403} (\bibinfo{year}{1972}).

\bibitem[{\citenamefont{Barash and Ginzburg}(1975)}]{barash1975a}
\bibinfo{author}{\bibfnamefont{Y.~S.} \bibnamefont{Barash}} \bibnamefont{and}
  \bibinfo{author}{\bibfnamefont{V.~L.} \bibnamefont{Ginzburg}},
  \bibinfo{journal}{Sov. Phys. Usp.} \textbf{\bibinfo{volume}{18}},
  \bibinfo{pages}{305} (\bibinfo{year}{1975}).

\bibitem[{\citenamefont{Narayanaswamy and Chen}(2010)}]{narayanaswamy2010a}
\bibinfo{author}{\bibfnamefont{A.}~\bibnamefont{Narayanaswamy}}
  \bibnamefont{and} \bibinfo{author}{\bibfnamefont{G.}~\bibnamefont{Chen}},
  \bibinfo{journal}{J. Quant. Spectrosc. Radiat. Transfer}
  \textbf{\bibinfo{volume}{111}}, \bibinfo{pages}{1877} (\bibinfo{year}{2010}),
  ISSN \bibinfo{issn}{0022-4073}.

\bibitem[{\citenamefont{Joulain et~al.}(2003)\citenamefont{Joulain, Carminati,
  Mulet, and Greffet}}]{joulain03a}
\bibinfo{author}{\bibfnamefont{K.}~\bibnamefont{Joulain}},
  \bibinfo{author}{\bibfnamefont{R.}~\bibnamefont{Carminati}},
  \bibinfo{author}{\bibfnamefont{J.-P.} \bibnamefont{Mulet}}, \bibnamefont{and}
  \bibinfo{author}{\bibfnamefont{J.-J.} \bibnamefont{Greffet}},
  \bibinfo{journal}{Phys. Rev. B} \textbf{\bibinfo{volume}{68}},
  \bibinfo{eid}{245405} (\bibinfo{year}{2003}).

\bibitem[{\citenamefont{Parsegian}(2005)}]{parsegian2005a}
\bibinfo{author}{\bibfnamefont{V.~A.} \bibnamefont{Parsegian}},
  \emph{\bibinfo{title}{Van der Waals Forces: A Handbook for Biologists,
  Chemists, Engineers, and Physicists}} (\bibinfo{publisher}{Cambridge
  University Press}, \bibinfo{year}{2005}).

\bibitem[{\citenamefont{Gingell and Parsegian}(1973)}]{gingell1973a}
\bibinfo{author}{\bibfnamefont{D.}~\bibnamefont{Gingell}} \bibnamefont{and}
  \bibinfo{author}{\bibfnamefont{V.}~\bibnamefont{Parsegian}},
  \bibinfo{journal}{J. Colloid Interface Sci.} \textbf{\bibinfo{volume}{44}},
  \bibinfo{pages}{456} (\bibinfo{year}{1973}), ISSN \bibinfo{issn}{0021-9797}.

\bibitem[{\citenamefont{Chew}(1995)}]{chew95a}
\bibinfo{author}{\bibfnamefont{W.~C.} \bibnamefont{Chew}},
  \emph{\bibinfo{title}{Waves and Fields in Inhomogeneous Media}}
  (\bibinfo{publisher}{IEEE Press}, \bibinfo{address}{Piscataway, NJ},
  \bibinfo{year}{1995}).

\bibitem[{\citenamefont{Pitaevskii}(to be
  published)}]{pitaevskiiCasimirPhysics}
\bibinfo{author}{\bibfnamefont{L.~P.} \bibnamefont{Pitaevskii}},
  \emph{\bibinfo{title}{Casimir Physics, Lecture Notes in Physics}}
  (\bibinfo{publisher}{Springer}, \bibinfo{year}{to be published}).

\end{thebibliography}

\end{document}

\title{Lifshitz theory of van der Waals pressure in dissipative media}% Force line breaks with \\
\section{Justification for existence of $\displaystyle \int\limits_{\infty}^{0} \partial T_{zz}^{avg}/\partial z_r {\rm d}z_v $}
Using Eq. 4 in the manuscript, we can see that the $zz$ component of the Maxwell stress tensor in the vacuum layer in Fig. 1b (in manuscript) is given by:
\begin{equation}
T_{zz}^{avg}(z_v) = \frac{k_BT}{\pi}\sum\limits_{n=0}^{\infty}{'}\int\limits_0^{\infty}  {\rm d}k_\rho  k_{zv} k_\rho \sum_{p=e,h} \frac{\widetilde{R}_{vL}^{(p)}\widetilde{R}_{vR}^{(p)}e^{-2k_{zv}z_v}}{1-\widetilde{R}_{vL}^{(p)}\widetilde{R}_{vR}^{(p)}e^{-2k_{zv}z_v}} 
\end{equation}
where $\widetilde{R}_{vL}$ and $\widetilde{R}_{vR}$ are generalized reflection coefficients that can be determined from the boundary conditions at each interface. $\widetilde{R}_{vL} \equiv \widetilde{R}_{vL}\left(z_1,\cdots,z_k\right)$ and $\widetilde{R}_{vR} \equiv \widetilde{R}_{vR}\left(z_{k+1},\cdots,z_N\right)$.
The work done in displacing the multilayer stack adjoining $R$ from $z_v = \infty$ to $z_v = \delta$ is given by:
\begin{equation}
\int\limits_{\infty}^{\delta} T_{zz}^{avg}(z_v) {\rm d}z_v =\frac{k_B T}{2\pi} \sum _{n=0}^{\infty}{'}\int\limits^{\infty}_{0} {\rm d}k_\rho k_\rho \sum_{p=e,h}\ln(1-\widetilde{R}_{vL}^{(p)}\widetilde{R}_{vR}^{(p)}e^{-2k_{zv}\delta})
\end{equation}

\begin{equation}
\label{eqn:limdiff}
\frac{\partial}{\partial z_r}\lim\limits_{\delta \rightarrow 0} \int\limits_{\infty}^{\delta} T_{zz}^{avg}(z_v) {\rm d}z_v =-\frac{k_B T}{2\pi} \sum _{n=0}^{\infty}{'}\lim\limits_{\delta \rightarrow 0} \int\limits^{\infty}_{0} {\rm d}k_\rho k_\rho \sum_{p=e,h} \frac{\displaystyle\frac{\partial}{\partial z_r}\left(\widetilde{R}_{vL}^{(p)}\widetilde{R}_{vR}^{(p)}\right)e^{-2k_{zv}\delta}}{1-\widetilde{R}_{vL}^{(p)}\widetilde{R}_{vR}^{(p)}e^{-2k_{zv}\delta}}
\end{equation}

$\displaystyle\frac{\partial}{\partial z_r}\left(\widetilde{R}_{vL}^{(p)}\widetilde{R}_{vR}^{(p)}\right) = \widetilde{R}_{vR}^{(p)} \frac{\partial \widetilde{R}_{vL}^{(p)}}{\partial z_r} $ or $ \displaystyle \widetilde{R}_{vL}^{(p)}\frac{\partial \widetilde{R}_{vR}^{(p)}}{\partial z_r} $, depending on whether $1\leq r \leq k$ or $k+1 \leq r \leq N$. The product $\widetilde{R}_{vL}^{(p)} \widetilde{R}_{vR}^{(p)}$ can be written in the form:
$\displaystyle \frac{A_1+A_2e^{-2k_{zr}z_r}}{A_3+A_4e^{-2k_{zr}z_r}}$, where $A_1,A_2,A_3,$ and $A_4$ are not functions of $z_r$. So, the partial differentiation $\displaystyle \frac{\partial}{\partial z_r}$ yields a function of the form 
$\displaystyle -2k_{zr}\frac{\left(A_3A_2-A_1A_4 \right)}{\left(A_3+A_4e^{-2k_{zr}z_r}\right)^2} e^{-2k_{zr}z_r}$. In general, we can write:
\begin{equation}
\frac{\partial}{\partial z_r}\left(\widetilde{R}_{vL}^{(p)}\widetilde{R}_{vR}^{(p)}\right)  = -2k_{zr}\frac{N}{D} e^{-2k_{zr}z_r}
\end{equation}
where $N \equiv A_3A_2-A_1A_4$ and $D \equiv \left(A_3+A_4e^{-2k_{zr}z_r}\right)^2$. Substituting the above relation in Eq. \ref{eqn:limdiff}, we get
\begin{equation}
\frac{\partial}{\partial z_r}\lim\limits_{\delta \rightarrow 0} \int\limits_{\infty}^{\delta} T_{zz}^{avg}(z_v) {\rm d}z_v =\frac{k_B T}{\pi} \sum _{n=0}^{\infty}{'}\lim\limits_{\delta \rightarrow 0} \int\limits^{\infty}_{0} {\rm d}k_\rho k_\rho k_{zr} \sum_{p=e,h} \frac{\displaystyle\frac{N}{D}e^{-2k_{zr}z_r}e^{-2k_{zv}\delta}}{1-\widetilde{R}_{vL}^{(p)}\widetilde{R}_{vR}^{(p)}e^{-2k_{zv}\delta}}
\end{equation}
Because of the presence of the $e^{-2k_{zr}z_r}$, we are justified in putting $\delta = 0$ to give:
\begin{equation}
\frac{\partial}{\partial z_r}\lim\limits_{\delta \rightarrow 0} \int\limits_{\infty}^{\delta} T_{zz}^{avg}(z_v) {\rm d}z_v =\frac{k_B T}{\pi} \sum _{n=0}^{\infty}{'}  \int\limits^{\infty}_{0} {\rm d}k_\rho k_\rho k_{zr} \sum_{p=e,h} \frac{\displaystyle\frac{N}{D}e^{-2k_{zr}z_r}}{1-\widetilde{R}_{vL}^{(p)}\widetilde{R}_{vR}^{(p)}} =  \int\limits_{\infty}^{0} \frac{\partial T_{zz}^{avg}}{\partial z_r} {\rm d}z_v 
\end{equation}
For the examples in our manuscripts, $\widetilde{R}_{vL}^{(p)}=R_{vL}^{(p)}$, $\widetilde{R}_{vR}^{(p)}=\displaystyle\frac{R_{vm}^{(p)}+R_{mR}^{(p)}e^{-2k_{zm}z_m}}{1+R_{vm}^{(p)}R_{mR}^{(p)}e^{-2k_{zm}z_m}}$, $N=R_{vL}^{(p)}R_{mR}^{(p)}(1-R_{vm}^{(p)2})$ and $D=(1+R_{vm}^{(p)}R_{mR}^{(p)}e^{-2k_{zm}z_m})^2$.

%I think we can write equations (7), (10) and (13) as follows, in the form of equation (1),
%
%\begin{equation}
%U_{mV}(z_m)=U_{VV}(z_m)+\lim\limits_{\delta\rightarrow0}\int\limits_{\infty}^{\delta} T_{zz}(z_m,z_v) {\rm d}z_v
%\end{equation}
%or
%\begin{equation}
%U_{mmV}(z_m)=U_{LV}(z_m)+U_{VmV}(z_m)+\lim\limits_{\delta\rightarrow0}\int\limits_{\infty}^{\delta} T_{zz}(z_m,z_v) {\rm d}z_v
%\end{equation}
%where, $U_{LV}(z_m)$, the vdW free energy of the semi--infinite planar medium, is 0.
%
%\begin{equation}
%U_{LV}(z_m)=U_{VV}(z_m)+\lim\limits_{\delta\rightarrow0}\int\limits_{\infty}^{\delta} T_{zz}(z_m,z_v) {\rm d}z_v
%\end{equation}
%or
%\begin{equation}
%U_{LmV}(z_m)=U_{LV}(z_m)+U_{VmV}(z_m)+\lim\limits_{\delta\rightarrow0}\int\limits_{\infty}^{\delta} T_{zz}(z_m,z_v) {\rm d}z_v
%\end{equation}
%
%\begin{equation}
%U_{LR}(z_m)=U_{VR}(z_m)+\lim\limits_{\delta\rightarrow0}\int\limits_{\infty}^{\delta} T_{zz}(z_m,z_v) {\rm d}z_v
%\end{equation}
%or
%\begin{equation}
%U_{LmR}(z_m)=U_{LV}(z_m)+U_{VmR}(z_m)+\lim\limits_{\delta\rightarrow0}\int\limits_{\infty}^{\delta} T_{zz}(z_m,z_v) {\rm d}z_v
%\end{equation}

\section{Derivation for Eq. 10 in Example 1}
\begin{equation}
\begin{split}
\label{ex1} &p_{VV}(z_m)=-\int\limits_{\infty}^{0} \frac{\partial T_{zz}^{avg}}{\partial z_m}{\rm d}z_v\\
&=-k_B T\sum _{n=0}^{\infty}{'}\int\limits_{\infty}^{0} \frac{\partial K_n}{\partial z_m}{\rm d}z_v\\
&\scriptstyle =- \frac{k_BT}{\pi}\sum _{n=0}^{\infty}{'} \int\limits_0^{\infty}\sum_{p=e,h}\frac{R_{vm}^{(p)}R_{mv}^{(p)}(1-R_{vm}^{(p)2})e^{-2k_{zm}z_m}}{[(1-R_{vm}^{(p)}R_{vm}^{(p)})-(R_{vm}^{(p)}R_{mv}^{(p)}+R_{mv}^{(p)}R_{mv}^{(p)})e^{-2k_{zm}z_m}](1+R_{vm}^{(p)}R_{mv}^{(p)}e^{-2k_{zm}z_m})} \scriptstyle k_{zm}k_\rho {\rm d}k_\rho\\
&=\frac{k_B T}{\pi}\sum _{n=0}^{\infty}{'}\int\limits_0^{\infty}\sum_{p=e,h}\frac{R_{mv}^{(p)2}e^{-2k_{zm}z_m}}{1-R_{mv}^{(p)2}e^{-2k_{zm}z_m}}k_{zm}k_\rho {\rm d}k_\rho\\
%&=\frac{k_BT}{\pi}\sum _{n=0}^{\infty}{'}  \int\limits_{\frac{\xi_n}{c}\sqrt{\varepsilon_m}}^{\infty} \sum_{p=e,h}\frac{R_{mv}^{(p)2}e^{-2k_{zm}z_m}}{1-R_{mv}^{(p)2}e^{-2k_{zm}z_m}} k_{zm}^2 dk_{zm}  \\
%&=\frac{k_BT}{\pi}\sum _{n=0}^{\infty}{'}  \frac{\xi_n^3}{c^3}\varepsilon_m^{3/2} \int\limits_{1}^{\infty} \sum_{p=e,h}\frac{R_{mv}^{(p)2}e^{-2k_{zm}z_m}}{1-R_{mv}^{(p)2}e^{-2k_{zm}z_m}} \frac{k_{zm}^2}{\frac{\xi_n^2}{c^2}\varepsilon_m} \frac{dk_{zm}}{\frac{\xi_n}{c}\sqrt{\varepsilon_m}}  \\
%&=\frac{k_BT}{\pi}\sum _{n=0}^{\infty}{'}  \frac{\xi_n^3}{c^3}\varepsilon_m^{3/2} \int\limits_{1}^{\infty} \sum_{p=e,h}\frac{R_{mv}^{(p)2}e^{-2p\varepsilon_n\sqrt{\varepsilon_m}z_m/c}}{1-R_{mv}^{(p)2}e^{-2p\varepsilon_n\sqrt{\varepsilon_m}z_m/c}} p^2 dp  \\
%&=\frac{k_BT}{\pi}\sum _{n=0}^{\infty}{'}  \frac{\xi_n^3}{c^3}\varepsilon_m^{3/2} \int\limits_{1}^{\infty} \left[(\overline{\triangle}^{-2}e^{2p\xi_n\sqrt{\varepsilon_m}z_m/c}-1)^{-1}+(\triangle^{-2}e^{2p\xi_n\sqrt{\varepsilon_m}z_m/c}-1)^{-1} \right] p^2dp  \\
%&=\frac{k_BT}{\pi c^3}\sum _{n=0}^{\infty}{'}  \varepsilon_m^{3/2} \xi_n^3\int\limits_{1}^{\infty} {\rm d}p p^2 \sum_{p=e,h} (R_{mv}^{(p)-2}e^{2p\xi_n\sqrt{\varepsilon_m}z_m/c}-1)^{-1}   \\
\end{split}
\end{equation}
%where,
%$p =\frac{k_{zm}}{\frac{\xi_n}{c}\sqrt{\varepsilon_m}} $.

\section{Derivation for Eq. 13 in Example 2}
\begin{equation}
\begin{split}
%\label{eqn:example2}
\label{ex2a} &p_{LV}(z_m)=p_{VV}(z_m)+\int\limits_{\infty}^{0}\frac{\partial T_{zz}^{avg}}{\partial z_m} {\rm d}z_v\\
&=p_{VV}(z_m)+k_B T\sum _{n=0}^{\infty}{'} \int\limits_{\infty}^{0} \frac{\partial K_n}{\partial z_m}{\rm d}z_v\\
%&=\frac{k_B T}{\pi}\sum _{n=0}^{\infty}{'}\int\limits_0^{\infty}\sum_{q=p,s}\frac{R_{mv}^{(q)2}e^{-2k_{zm}z_m}}{1-R_{mv}^{(q)2}e^{-2k_{zm}z_m}}k_{zm}k_\rho {\rm d}k_\rho+\\
%&\frac{k_BT}{\pi}\sum _{n=0}^{\infty}{'} \int\limits_0^{\infty}\sum_{q=p,s}\frac{R_{vL}^{(q)}R_{mv}^{(q)}(1-R_{vm}^{(q)2})e^{-2k_{zm}z_m}}{[(1-R_{vL}^{(q)}R_{vm}^{(q)})-(R_{vL}^{(q)}R_{mv}^{(q)}+R_{mv}^{(q)}R_{mv}^{(q)})e^{-2k_{zm}z_m}](1+R_{vm}^{(q)}R_{mv}^{(q)}e^{-2k_{zm}z_m})}k_{zm}k_\rho {\rm d}k_\rho\\
%&=\frac{k_B T}{\pi}\sum _{n=0}^{\infty}{'}\int\limits_0^{\infty}\sum_{p=e,h}\left( \frac{R_{mv}^{(p)2}e^{-2k_{zm}z_m}}{1-R_{mv}^{(p)2}e^{-2k_{zm}z_m}}+\frac{R_{VL}^{(p)}R_{mv}^{(p)}(1-R_{Vm}^{(p)2})e^{-2k_{zm}z_m}}{[(1-R_{VL}^{(p)}R_{Vm}^{(p)})-(R_{VL}^{(p)}R_{mv}^{(p)}+R_{mv}^{(p)}R_{mv}^{(p)})e^{-2k_{zm}z_m}](1+R_{Vm}^{(p)}R_{mv}^{(p)}e^{-2k_{zm}z_m})} \right) k_{zm}k_\rho {\rm d}k_\rho\\
&\scriptstyle =\frac{k_B T}{\pi}\sum _{n=0}^{\infty}{'}\int\limits_0^{\infty}\sum_{p=e,h} \frac{(R_{vL}^{(p)}+R_{mv}^{(p)})(1-R_{vm}^{(p)2}e^{-2k_{zm}z_m})R_{mv}^{(p)}e^{-2k_{zm}z_m}}{[(1-R_{vL}^{(p)}R_{vm}^{(p)})-(R_{vL}^{(p)}R_{mv}^{(p)}+R_{mv}^{(p)}R_{mv}^{(p)})e^{-2k_{zm}z_m}](1-R_{mv}^{(p)2}e^{-2k_{zm}z_m})}  k_{zm}k_\rho {\rm d}k_\rho\\
&=\frac{k_B T}{\pi}\sum _{n=0}^{\infty}{'}\int\limits_0^{\infty}\sum_{p=e,h} \displaystyle \frac{(R_{mv}^{(p)}+R_{vL}^{(p)})R_{mv}^{(p)}e^{-2k_{zm}z_m}}{(1+R_{Lv}^{(p)}R_{vm}^{(p)})-(R_{vL}^{(p)}R_{mv}^{(p)}+R_{mv}^{(p)}R_{mv}^{(p)})e^{-2k_{zm}z_m}}  k_{zm}k_\rho {\rm d}k_\rho\\
&=\frac{k_B T}{\pi}\sum _{n=0}^{\infty}{'}\int\limits_0^{\infty}\sum_{p=e,h} \frac{\displaystyle \frac{R_{mv}^{(p)}+R_{vL}^{(p)}}{1+R_{mv}^{(p)}R_{vL}^{(p)}} R_{mv}^{(p)}e^{-2k_{zm}z_m}}{1-\displaystyle \frac{R_{mv}^{(p)}+R_{vL}^{(p)}}{1+R_{mv}^{(p)}R_{vL}^{(p)}}R_{mv}^{(p)}e^{-2k_{zm}z_m}}  k_{zm}k_\rho {\rm d}k_\rho\\
&=\frac{k_B T}{\pi}\sum _{n=0}^{\infty}{'}\int\limits_0^{\infty}\sum_{p=e,h} \frac{R_{mL}^{(p)}R_{mv}^{(p)}e^{-2k_{zm}z_m}}{1-R_{mL}^{(p)}R_{mv}^{(p)}e^{-2k_{zm}z_m}} k_{zm}k_\rho {\rm d}k_\rho\\
%&=\frac{k_BT}{\pi c^3}\sum _{n=0}^{\infty}{'}  \varepsilon_m^{3/2} \xi_n^3\int\limits_{1}^{\infty} p^2\left[(\overline{\triangle}^{-2}e^{2p\xi_n\sqrt{\varepsilon_m}z_m/c}-1)^{-1}+(\triangle^{-2}e^{2p\xi_n\sqrt{\varepsilon_m}z_m/c}-1)^{-1} \right] dp  \\
\end{split}
\end{equation}
where we have used the following property of reflection coefficients: %\textbf{Put the fractions like $\frac{k_{zm}}{\mu_m}$. If they appear too small, use $\backslash$displaystyle before the frac command to get $\displaystyle \frac{k_{zm}}{\mu_m}$}
\begin{equation}
\begin{split}
\frac{R_{mv}^{(e)}+R_{vL}^{(e)}}{1+R_{mv}^{(e)}R_{vL}^{(e)}}&=\frac{\left(\displaystyle \frac{k_{zm}}{\mu_m}-\displaystyle \frac{k_{zv}}{\mu_v}\right)\left(\displaystyle \frac{k_{zv}}{\mu_v}+\displaystyle \frac{k_{zL}}{\mu_L}\right)+\left(\displaystyle \frac{k_{zm}}{\mu_m}+\displaystyle \frac{k_{zv}}{\mu_v}\right)\left(\displaystyle \frac{k_{zv}}{\mu_v}-\displaystyle \frac{k_{zL}}{\mu_L}\right)}{\left(\displaystyle \frac{k_{zm}}{\mu_m}+\displaystyle \frac{k_{zv}}{\mu_v}\right)\left(\displaystyle \frac{k_{zv}}{\mu_v}+\displaystyle \frac{k_{zL}}{\mu_L}\right)+\left(\displaystyle \frac{k_{zm}}{\mu_m}-\displaystyle \frac{k_{zv}}{\mu_v}\right)\left(\displaystyle \frac{k_{zv}}{\mu_v}-\displaystyle \frac{k_{zL}}{\mu_L}\right)}\\
&=\frac{\displaystyle \frac{k_{zv}}{\mu_v}\left(\displaystyle \frac{k_{zm}}{\mu_m}-\displaystyle \frac{k_{zL}}{\mu_L}\right)}{\displaystyle \frac{k_{zv}}{\mu_v}\left(\displaystyle \frac{k_{zm}}{\mu_m}+\displaystyle \frac{k_{zL}}{\mu_L}\right)}=\frac{\displaystyle \frac{k_{zm}}{\mu_m}-\displaystyle \frac{k_{zL}}{\mu_L}}{\displaystyle \frac{k_{zm}}{\mu_m}+\displaystyle \frac{k_{zL}}{\mu_L}}=R_{mL}^{(e)}\\
\end{split}
\end{equation}
to go from the last but one to the last line of Eq. \ref{ex2a}. Similarly, $\displaystyle \frac{R_{mv}^{(h)}+R_{vL}^{(h)}}{1+R_{mv}^{(h)}R_{vL}^{(h)}}=R_{mL}^{(h)}$.

\section{Derivation for Eq. 15 in Example 3}
%Use the appropriate $\backslash$scriptstyle and $\backslash$displaystyle commands to make the derivation below more readable.
\begin{equation}
\begin{split}
\label{ex3} &p_{LR}(z_m)=p_{VR}(z_m)+\int\limits_{\infty}^{0} \frac{\partial T_{zz}^{avg}}{\partial z_m} {\rm d}z_v\\
&=p_{VR}(z_m)+k_B T\sum _{n=0}^{\infty}{'} \lim\limits_{\delta\rightarrow0}\int\limits_{\infty}^{\delta} \frac{\partial K_n}{\partial z_m}{\rm d}z_v\\
%&=\frac{k_B T}{\pi}\sum _{n=0}^{\infty}{'}\int\limits_0^{\infty}\sum_{q=p,s} \frac{R_{mv}^{(q)}R_{mR}^{(q)}e^{-2k_{zm}z_m}}{1-R_{mv}^{(q)}R_{mR}^{(q)}e^{-2k_{zm}z_m}} k_{zm}k_\rho {\rm d}k_\rho+\\
%&\frac{k_BT}{\pi}\sum _{n=0}^{\infty}{'} \int\limits_0^{\infty}\sum_{q=p,s}\frac{R_{vL}^{(q)}R_{mR}^{(q)}(1-R_{vm}^{(q)2})e^{-2k_{zm}z_m}}{[(1-R_{vL}^{(q)}R_{vm}^{(q)})-(R_{vL}^{(q)}R_{mR}^{(q)}+R_{mv}^{(q)}R_{mR}^{(q)})e^{-2k_{zm}z_m}](1+R_{vm}^{(q)}R_{mR}^{(q)}e^{-2k_{zm}z_m})}k_{zm}k_\rho {\rm d}k_\rho\\
%&=\frac{k_B T}{\pi}\sum _{n=0}^{\infty}{'}\int\limits_0^{\infty}\sum_{p=e,h}\left( \frac{R_{mv}^{(p)}R_{mR}^{(p)}e^{-2k_{zm}z_m}}{1-R_{mv}^{(p)}R_{mR}^{(p)}e^{-2k_{zm}z_m}}+\frac{R_{vL}^{(p)}R_{mv}^{(p)}(1-R_{Vm}^{(p)2})e^{-2k_{zm}z_m}}{[(1-R_{vL}^{(p)}R_{Vm}^{(p)})-(R_{vL}^{(p)}R_{mv}^{(p)}+R_{mv}^{(p)}R_{mv}^{(p)})e^{-2k_{zm}z_m}](1+R_{Vm}^{(p)}R_{mv}^{(p)}e^{-2k_{zm}z_m})} \right) k_{zm}k_\rho {\rm d}k_\rho\\
&\scriptstyle=\frac{k_B T}{\pi}\sum _{n=0}^{\infty}{'}\int\limits_0^{\infty}\sum_{p=e,h} \frac{(R_{vL}^{(p)}+R_{mv}^{(p)})(1+R_{vm}^{(p)}R_{mR}^{(p)}e^{-2k_{zm}z_m})R_{mR}^{(p)}e^{-2k_{zm}z_m}}{[(1-R_{vL}^{(p)}R_{vm}^{(p)})-(R_{vL}^{(p)}R_{mR}^{(p)}+R_{mv}^{(p)}R_{mR}^{(p)})e^{-2k_{zm}z_m}](1+R_{vm}^{(p)}R_{mR}^{(p)}e^{-2k_{zm}z_m})}  k_{zm}k_\rho {\rm d}k_\rho\\
&=\frac{k_B T}{\pi}\sum _{n=0}^{\infty}{'}\int\limits_0^{\infty}\sum_{p=e,h} \frac{(R_{vL}^{(p)}+R_{mv}^{(p)})R_{mR}^{(p)}e^{-2k_{zm}z_m}}{(1-R_{vL}^{(p)}R_{vm}^{(p)})-(R_{vL}^{(p)}R_{mR}^{(p)}+R_{mv}^{(p)}R_{mR}^{(p)})e^{-2k_{zm}z_m}}  k_{zm}k_\rho {\rm d}k_\rho\\
&=\frac{k_B T}{\pi}\sum _{n=0}^{\infty}{'}\int\limits_0^{\infty}\sum_{p=e,h} \frac{\displaystyle{\frac{R_{vL}^{(p)}+R_{mv}^{(p)}}{1+R_{vL}^{(p)}R_{mv}^{(p)}}} R_{mR}^{(p)}e^{-2k_{zm}z_m}}{\displaystyle{\frac{1+R_{vL}^{(p)}R_{mv}^{(p)}}{{1+R_{vL}^{(p)}R_{mv}^{(p)}}}}-\displaystyle{\frac{R_{vL}^{(p)}+R_{mv}^{(p)}}{1+R_{vL}^{(p)}R_{mv}^{(p)}}}R_{mR}^{(p)}e^{-2k_{zm}z_m}}  k_{zm}k_\rho {\rm d}k_\rho\\
&=\frac{k_B T}{\pi}\sum _{n=0}^{\infty}{'}\int\limits_0^{\infty}\sum_{p=e,h} \frac{R_{mL}^{(p)}R_{mR}^{(p)}e^{-2k_{zm}z_m}}{1-R_{mL}^{(p)}R_{mR}^{(p)}e^{-2k_{zm}z_m}} k_{zm}k_\rho {\rm d}k_\rho\\
\end{split}
\end{equation}

\section{Derivation for Eq. 16}
\begin{equation}
\begin{split}
&p_{LR}(z_m)=\frac{k_B T}{\pi}\sum _{n=0}^{\infty}{'}\int\limits_0^{\infty}\sum_{p=e,h} \frac{R_{mL}^{(p)}R_{mR}^{(p)}e^{-2k_{zm}z_m}}{1-R_{mL}^{(p)}R_{mR}^{(p)}e^{-2k_{zm}z_m}} k_{zm}k_\rho {\rm d}k_\rho\\
&=\frac{k_B T}{\pi}\sum _{n=0}^{\infty}{'}\int\limits_{\frac{\xi_n}{c}\sqrt{\varepsilon_m}}^{\infty}\sum_{p=e,h} \frac{R_{mL}^{(p)}R_{mR}^{(p)}e^{-2k_{zm}z_m}}{1-R_{mL}^{(p)}R_{mR}^{(p)}e^{-2k_{zm}z_m}} k_{zm}^2 {\rm d}k_{zm}\\
&=\frac{k_B T}{\pi}\sum _{n=0}^{\infty}{'}\frac{\xi_n^3}{c^3}\varepsilon_m^{3/2} \int\limits_1^{\infty}\sum_{p=e,h} \frac{R_{mL}^{(p)}R_{mR}^{(p)}e^{-2k_{zm}z_m}}{1-R_{mL}^{(p)}R_{mR}^{(p)}e^{-2k_{zm}z_m}} \frac{k_{zm}^2}{\frac{\xi_n^2}{c^2}\varepsilon_m} \frac{{\rm d}k_{zm}}{\frac{\xi_n}{c}\sqrt{\varepsilon_m}} \\
&=\frac{k_B T}{\pi}\sum _{n=0}^{\infty}{'}\frac{\xi_n^3}{c^3}\varepsilon_m^{3/2} \int\limits_1^{\infty}\sum_{p=e,h} \frac{R_{mL}^{(p)}R_{mR}^{(p)}e^{-2q\xi_n\sqrt{\varepsilon_m} z_m/c}}{1-R_{mL}^{(p)}R_{mR}^{(p)}e^{-2q\xi_n\sqrt{\varepsilon_m} z_m/c}}q^2{\rm d}q\\
&=\frac{k_B T}{\pi c^3}\sum _{n=0}^{\infty}{'}\varepsilon_m^{3/2}\xi_n^3 \int\limits_1^{\infty}\sum_{p=e,h} \left(R_{mL}^{(p)-1}R_{mR}^{(p)-1}e^{2q\xi_n\sqrt{\varepsilon_m} z_m/c}-1\right)^{-1}q^2{\rm d}q\\
\end{split}
\end{equation}
where, $q=k_{zm}/(\xi_n\sqrt{\varepsilon_m(i\xi_n)}/c)$.